\documentclass[10pt,journal,compsoc]{IEEEtran}

\ifCLASSOPTIONcompsoc
  \usepackage[nocompress]{cite}
\else
  \usepackage{cite}
\fi

\usepackage{threeparttable}
\usepackage{cite}
\usepackage{amsmath,amssymb,amsfonts}
\usepackage{algorithmic}
\usepackage{graphicx}
\usepackage{textcomp}
\usepackage{indentfirst}
\usepackage{csquotes}
\usepackage{multirow}
\usepackage{tabularx}
\usepackage{float}
\usepackage{color}
\usepackage{comment}
\usepackage[ruled,linesnumbered]{algorithm2e}

\usepackage{url}

\usepackage{multirow}
\newcommand{\squishlist}{
   \begin{list}{$\bullet$}
    { \setlength{\itemsep}{0pt}      \setlength{\parsep}{0pt}
      \setlength{\topsep}{3pt}       \setlength{\partopsep}{0pt}
      \setlength{\listparindent}{-2pt}
      \setlength{\itemindent}{-5pt}
      \setlength{\leftmargin}{1em} \setlength{\labelwidth}{0em}
      \setlength{\labelsep}{0.5em} } }

\newcommand{\squishend}{
    \end{list}  }
    
\usepackage{amssymb}
\usepackage{pifont}

\usepackage[ruled,linesnumbered]{algorithm2e}
\usepackage{pifont}
\usepackage{xcolor}
\usepackage{booktabs}
\usepackage{graphicx}  
\usepackage{stfloats}

\usepackage{tikz}
\usepackage{enumitem}
\usetikzlibrary{shapes,snakes}
\usepackage{multirow}
\usepackage{booktabs}
\usepackage{comment}
\usepackage{color}

\newcommand*\blackcircledempty[1]{\tikz[baseline=(char.base)]{
        \node[shape=circle, text={rgb,255:red,0;green,0;blue,0}, font=\small, draw={rgb,255:red,0;green,0;blue,0},inner sep=0.5pt] (char) {#1};}}

\usepackage[explicit]{titlesec}	
\titlespacing*{\section}{0pt}{1.2ex plus .0ex minus .0ex}{.3ex plus .0ex}
\titlespacing*{\subsection}{0pt}{1.2ex plus .0ex minus .0ex}{.3ex plus .0ex}

\begin{document}


\title{Characterizing and Understanding \\ HGNNs on GPUs}


\author{
Mingyu~Yan, 
Mo~Zou, 
Xiaocheng~Yang, 
Wenming~Li, \\
Xiaochun~Ye, 
Dongrui~Fan,~\IEEEmembership{Senior~Member,~IEEE},
and~Yuan~Xie,~\IEEEmembership{Fellow,~IEEE}

\IEEEcompsocitemizethanks{
\IEEEcompsocthanksitem 
M.~Yan, M.~Zou, X.~Yang, W.~Li, X.~Ye, and D.~Fan are with the Institute of Computing Technology, Chinese Academy of Sciences, Beijing, China. M.~Zou is also with University of Chinese Academy of Sciences, Beijing, China. E-mail: yanmingyu, zoumo, liwenming, yexiaochun, fandr@ict.ac.cn. Y. Xie is with University of California, Santa Barbara, California, USA. Email: yuanxie@ece.ucsb.edu.

\IEEEcompsocthanksitem 
This work was supported by CAS Project for Young Scientists in Basic Research (Grant No. YSBR-029), and CAS Project for Youth Innovation Promotion Association. Corresponding author is Mo Zhou.

}
}

\markboth{To Appear in IEEE Computer Architecture Letters}%
{Shell \MakeLowercase{\textit{et al.}}: Bare Demo of IEEEtran.cls for Computer Society Journals}

\IEEEtitleabstractindextext{%
\begin{abstract}
Heterogeneous graph neural networks (HGNNs) deliver powerful capacity in heterogeneous graph representation learning. The execution of HGNNs is usually accelerated by GPUs. Therefore, characterizing and understanding the execution pattern of HGNNs on GPUs is important for both software and hardware optimizations. Unfortunately, there is no detailed characterization effort of HGNN workloads on GPUs. In this paper, we characterize HGNN workloads at inference phase and explore the execution of HGNNs on GPU, to disclose the execution semantic and execution pattern of HGNNs. Given the characterization and exploration, we propose several useful guidelines for both software and hardware optimizations for the efficient execution of HGNNs on GPUs.
\end{abstract}

}

\maketitle

\IEEEdisplaynontitleabstractindextext

\IEEEpeerreviewmaketitle

\IEEEraisesectionheading{
\section{Introduction}\label{sec:introduction}}

\IEEEPARstart{I}{n} recent years, heterogeneous graph neural networks (HGNNs) have attracted much attention in graph representation learning, as they deliver powerful capacity to capture the rich structural and semantic information from the heterogeneous graph (HG). 
Unlike graph neural networks (GNNs) which learn from the homogeneous graph constituted by one type of node and edge, HGNNs learn from the HG consisted of multi-types of nodes and edges. 
HGNNs are accelerated by GPUs to achieve a considerable performance. Therefore, disclosing the execution pattern and execution semantic of HGNNs on GPUs is important for both software and hardware optimizations for HGNNs.


The complex structure and semantic of HGs offer a more realistic application scenario for HGNNs, but also impose great challenges on the optimization of HGNNs.
Compared to GNNs, HGNNs aggregate not only the structural information but also semantic information due to the multi-types of nodes and edges in HGs. 
To this end, most of prevalent HGNNs usually contain four major execution stages: \textit{Subgraph Build}, \textit{Feature Projection}, \textit{Neighbor Aggregation}, and \textit{Semantic Aggregation}, making the execution patterns differ from traditional workloads.
Therefore, observations and conclusions in previous characterization efforts~\cite{gnn_characterization,hygcn,zhang2020architectural} for GNNs cannot be directly inferred in HGNNs.

To disclose the execution pattern and execution semantic of HGNNs, we characterize the inference phase of three prevalent HGNNs with three well-known HGs on GPU. The key observations and insights are summarized below.

\begin{itemize}[leftmargin=*]
    \item \textbf{Overall Analysis}: 
    1) \textit{Neighbor Aggregation} stage dominates most time of HGNNs; 
    2) Most time is consumed by the kernel with reduction-tree-based computational graph. 
    \item \textbf{Detail Analysis for Each Stage}: 
    1) \textit{Feature Projection} stage is dominated by the execution of dense-dense matrix multiplication, primarily facing compute bound;
    2) \textit{Neighbor Aggregation} stage is dominated by the execution of graph-topology-based and element-wise operations, primarily facing memory bound and exhibiting irregular memory access pattern;
    3) \textit{Semantic Aggregation} stage is dominated by the execution of dense-dense matrix multiplication, element-wise operation, and data rearrangement operation, primarily facing memory bound first and then compute bound;
    4) The data rearrangement in \textit{Semantic Aggregation} stage is expensive. 
    
    \item \textbf{Comparison}: 
    1) Except for increasing with the average number of neighbors as in \textit{Aggregation} stage in GNNs, the execution time of \textit{Neighbor Aggregation} stage in HGNNs increases further as the number of metapaths increases;
    2) Inter-subgraph parallelism exists in \textit{Neighbor Aggregation }stage in HGNNs;  
    3) A barrier exists between \textit{Neighbor Aggregation} and \textit{Semantic Aggregation} stages of HGNNs.
    \item \textbf{Exploration}: 
    1) The sparsity of subgraph decreases as the length of metapath increases;
    2) The total execution time increases as the number of metapaths increases.
\end{itemize}

\section{Background}\label{sec:background}


\textbf{HGs.} HGs consist of multiple types of nodes and/or multiple types of edges. Each two nodes in a HG can be connected via different semantic paths, which are called metapaths. A metapath is defined as a path in the form of {\small$  t_{1} \stackrel{r_{1}}{\longrightarrow} t_{2} 
\cdots \stackrel{r_{l}}{\longrightarrow} t_{1+1} $} (abbr. as $t_1 t_2 \cdots t_{l+1}$) with node types $t_1, 
\cdots, t_{l+1}$ and edge types $r_1, 
\cdots, r_{l}$. Given a node $v$ and a metapath $P$ in a HG, the metapath-based neighbors of $v$ are defined as the set of nodes which connect with $v$ via $P$.

\begin{figure*}[!t]
        \vspace{-5pt}
 	    \centering
    	\includegraphics[width = \linewidth]{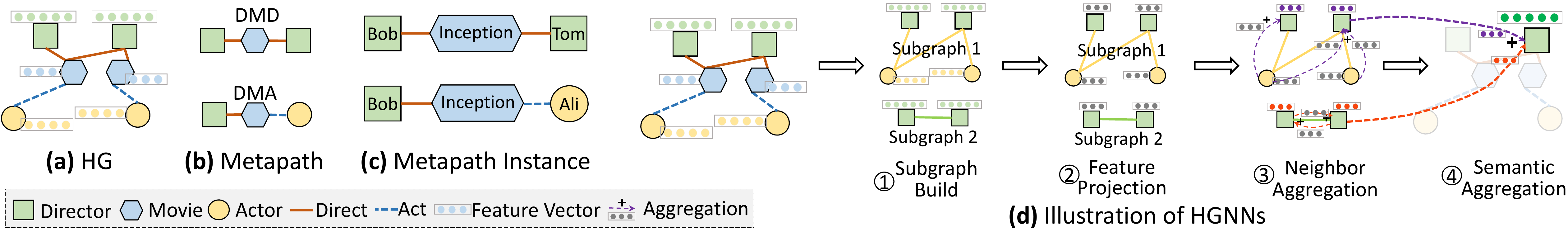}
        \vspace{-25pt}
    	\caption{Illustrations of (a) HG, (b) metapath, (c) metapath instance, and (d) HGNN.}
    	\label{fig:HG}
        \vspace{-15pt}
\end{figure*}

Fig.~\ref{fig:HG}(a) illustrates an example of a HG (IMDB dataset).
IMDB dataset consists of three types of nodes (Director (D), Movie (M), and Actor (A)) and two types of edges corresponding to two relations (Direct relation between movie and director, Act relation between actor and movie).
Considering a metapath DMD in Fig.~\ref{fig:HG}(b),  
\textit{Bob}-\textit{Inception}-\textit{Tom} is a metapath instance and Director \textit{Bob} is a metapath-based neighbor of Director \textit{Tom}, indicating the co-director relationship in movie \textit{Inception}, as shown in Fig.~\ref{fig:HG}(c).

\textbf{HGNNs.} To capture structural and semantic information, the executions of HGNNs can be generally categorized into four primary stages as shown in Fig.~\ref{fig:HG}(d):
%
\blackcircledempty{1} \textit{\textbf{Subgraph Build}} splits a HG into multiple subgraphs using relation walk or metapath walk. 
\blackcircledempty{2} \textit{\textbf{Feature Projection (FP)}} projects feature vectors of different types of nodes into one latent vector space. A type-specific linear transformation for each type of nodes is performed for the projection. 
\blackcircledempty{3} \textit{\textbf{Neighbor Aggregation (NA)}} aggregates feature vectors of neighbor nodes for each node within each subgraph to capture the structural information. 
\blackcircledempty{4} \textit{\textbf{Semantic Aggregation (SA)}} aggregates semantic information revealed by all metapaths, i.e., aggregates the results of \textit{Neighbor Aggregation} stage across different subgraphs for each node, with the consideration of the importance of different metapaths. In this work, we focus on metapath-based HGNNs and summarize three prevalent HGNNs briefly in Table~\ref{tab:hgnn_models}.

\textbf{Differences between HGNNs and GNNs.} Traditional GNNs usually consist of two stages: \textit{Aggregation} stage aggregates the feature vectors from neighbor nodes; \textit{Combination} stage updates feature vectors of each node using the aggregating results. 
The major differences between HGNNs and GNNs are as follows:
\squishlist
    \item \textit{Heterogeneous Input versus Homogeneous Input}. %
    There exists only one type of node with the same dimension feature vectors in the input homogeneous graph of GNNs. 
    But the multi-types of nodes in HGs have different attributes which need an extra \textit{Feature Projection} stage in HGNNs to project raw feature vectors with various dimensions into ones with the same dimension in a latent vector space.
    
    \item \textit{Multiple Semantics versus Single Semantic}. 
    There exists only one type of relation between each pair of nodes in homogeneous graphs, which implies that only one type of semantic information can be captured in GNNs. 
    On the contrary, each two nodes can be connected via different metapaths in HGs, which reveals that multiple semantic information can be captured in HGNNs. An extra \textit{Semantic Aggregation} stage is used to capture these semantic information in HGNNs. 

    \item \textit{Two-stage Aggregation versus One-stage Aggregation}. 
    Unlike GNNs which only aggregate once for the neighbor aggregation, HGNNs need both neighbor and semantic aggregations, while the former aggregates neighbor information from intra-metapaths and the latter aggregates semantic information from inter-metapaths.
    
\squishend


\section{Evaluation Setup}\label{sec:evaluation setup}
\textbf{Benchmark.} 
Table \ref{tab:hgnn_models} and \ref{tab:datasets} provide details of the benchmark HGNN models and HG datasets used in our experiments. 
The HAN, RGCN, and MAGNN are implemented based on the state-of-the-art framework DGL 0.7.2~\cite{DGL}. See their source codes in OpenHGNN 0.2.0 \footnote{https://github.com/BUPT-GAMMA/OpenHGNN} and MAGNN \footnote{https://github.com/cynricfu/MAGNN}. 

\begin{table}[!t]
    \setlength\tabcolsep{3pt}%
    \vspace{-15pt}
    \caption{Primary operations of four stages across three HGNN models.}  \label{tab:hgnn_models}
	\vspace{-10pt}
	\centering
    \renewcommand\arraystretch{1.3}
    \resizebox{0.49\textwidth}{!}{
    \begin{tabular}{c|c|c|c|c}
    \toprule
          & \blackcircledempty{1} & \blackcircledempty{2} & \blackcircledempty{3} & \blackcircledempty{4} \\
         \midrule
         R-GCN~\cite{RGCN} & Relation Walk & Linear Transformation  & Mean & Sum  \\
         
         HAN~\cite{HAN} & Metapath Walk & Linear Transformation  & GAT & Attention Sum\\
         
         MAGNN~\cite{MAGNN} & Metapath Walk & Linear Transformation  & GAT & Attention Sum\\
         \bottomrule
    \end{tabular}
}
\vspace{-12pt}
\end{table}

\begin{table}[!t]
	\caption{Information of HG datasets and Reddit dataset.}\label{tab:datasets}
	\vspace{-10pt}
	\centering
    \renewcommand\arraystretch{0.8}
    \resizebox{0.45\textwidth}{!}{
\begin{tabular}{cccc}
\toprule
Dataset                                                                & Node               & Feature Dimension               & Relation   \\ \midrule

\multirow{4}{*}{\begin{tabular}[c]{@{}c@{}}IMDB\\ (IM)\end{tabular}}   & movie (M): 4278    & M: 3066                      & A-M: 12828 \\
                                                                       & director (D): 2081 & D: 2081                      & D-M: 4278  \\
                                                                       & actor (A): 5257    & A: 5257                      & M-A: 12828 \\
                                                                       &                    &                              & M-D: 4278  \\ \midrule
\multirow{4}{*}{\begin{tabular}[c]{@{}c@{}}ACM\\ (AC)\end{tabular}}    & author (A): 5912   & A: 1902                      & P-A: 9936  \\
                                                                       & paper (P): 3025    & P: 1902                      & P-S: 3025  \\
                                                                       & subject (S): 57    & S: 1902                      & A-P: 9936  \\
                                                                       &                    &                              & P-S: 3025  \\ \midrule
\multirow{4}{*}{\begin{tabular}[c]{@{}c@{}}DBLP\\ (DB)\end{tabular}}   & author (A): 4057   & A: 334                       & P-A: 19645 \\
                                                                       & paper (P): 14328   & P: 14328                     & P-T: 85810 \\
                                                                       & Term (T): 7723     & T: 7723                      & P-V: 14328 \\
                                                                       & Venue (V): 20      & V: 20                        &            \\ \midrule
Reddit (RD)      & 232965 &  602  &  114615892 \\
\bottomrule
\end{tabular}
}
\vspace{-15pt}
\end{table}

\textbf{Profiling Platform.} 
All the workloads are profiled on a NVIDIA GPU T4 using NVIDIA Nsight System and Nsight Compute command lines. 



\section{Observation and Analysis}\label{sec:observation and analysis}
\subsection{Overview of Profile}\label{subsec:overview of profile}
\begin{figure}[!t]
        \vspace{-5pt}
 	    \centering
    	\includegraphics[width = 0.89\linewidth]{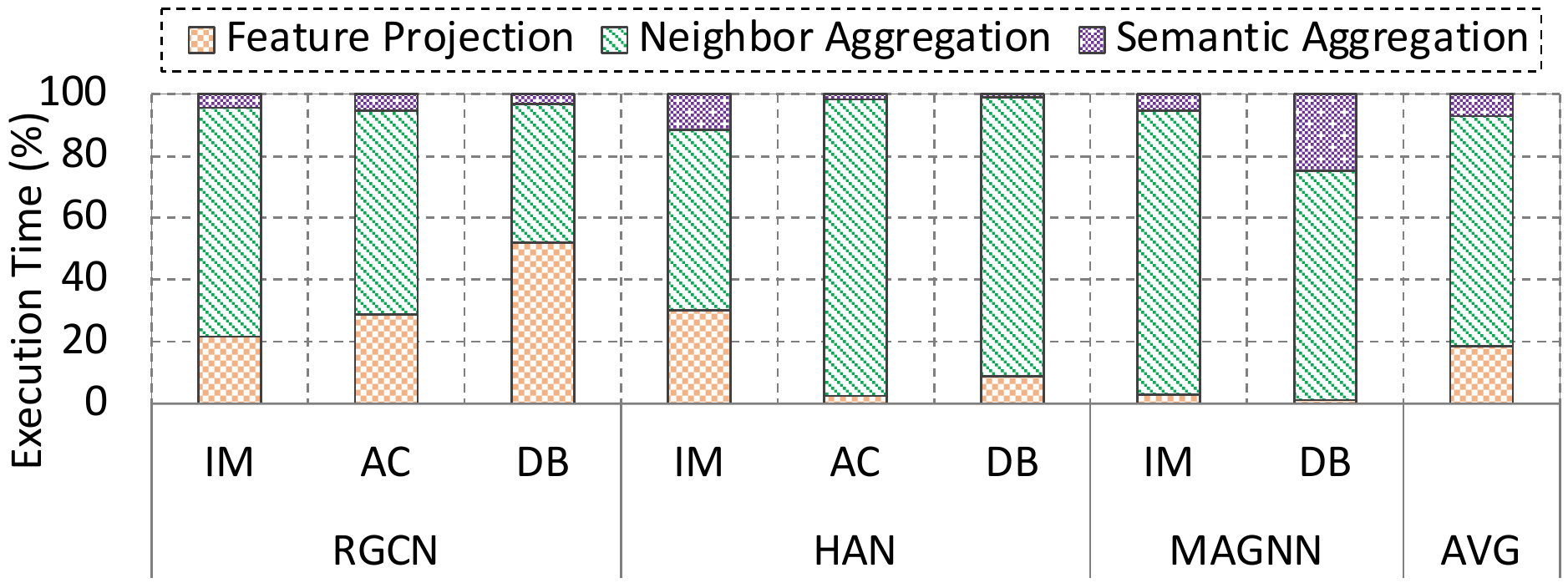}
    	\vspace{-15pt}
    	\caption{Execution time breakdown.}
    	\label{fig:execution_breakdown}
\end{figure}

\textbf{Execution Time Breakdown.} Fig.~\ref{fig:execution_breakdown} shows the execution time breakdown of inference phase. We omit \textit{Subgraph Build} stage since it is executed in CPU before inference phase.

\textit{Neighbor Aggregation stage dominates most execution time of HGNNs.} 
Fig.~\ref{fig:execution_breakdown} shows that \textit{Feature Projection}, \textit{Neighbor Aggregation}, and \textit{Semantic Aggregation} stages take 19\%, 74\%, and 7\% execution time averaging across different models and datasets, respectively. \textit{Neighbor Aggregation} stage takes most time. This is because for each subgraph, neighboring feature vectors of each node in the corresponding subgraph all need to be aggregated, which is time-consuming.



\begin{figure}[!t]
        \vspace{-10pt}
 	    \centering
    	\includegraphics[width = 0.89\linewidth]{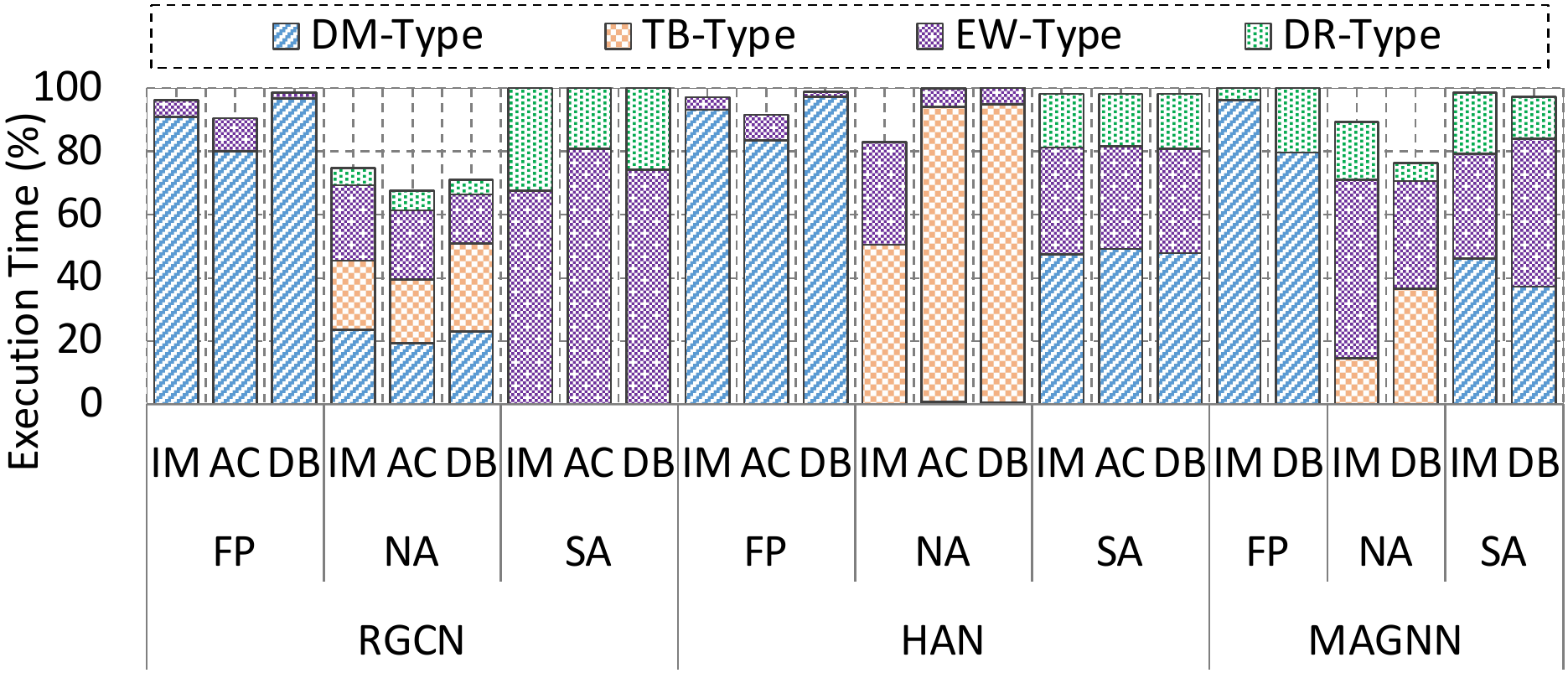}
    	\vspace{-15pt}
    	\caption{Execution time breakdown on different types of CUDA kernels.}
    	\label{fig:stage_breakdown}
    	\vspace{-15pt}
\end{figure}

\textbf{Execution time breakdown on different types of CUDA kernels.}
To further disclose the overall execution of HGNNs, Fig. \ref{fig:stage_breakdown} shows four major kernel types that occupy most execution time on the three stages across three models and three datasets.
These kernel types include dense-dense matrix multiplication (DeMM) kernel (DM-Type), topology-based matrix operation kernel (TB-Type), element-wise compute kernel (EW-Type), and data rearrangement kernel (DR-Type).
The DM-Type kernels perform DeMM such as the \textit{sgemm}, generally exhibiting compute bound due to their regular execution pattern and high compute-to-memory access ratio.
The TB-Type kernels perform the compute operations based on the irregular topology of graph such as the \textit{SpMMCsr} and \textit{SDDMMCoo (SDDMM, sampled dense-dense matrix multiplication)}, generally exhibiting memory bound derived from the irregular execution pattern caused by the irregular neighbor connection pattern in graph.
The EW-Type kernels perform element-wise compute operations on a set of vectors or a matrix such as the \textit{unrolled\_elementwise\_kernel (uEleWise)}, \textit{vectorized\_elementwise\_kernel (vEleWise)}, and \textit{reduce\_kernel (Reduce)}, generally exhibiting memory bound due to low compute-to-communication ratio.
The DR-Type kernels perform data rearrangement on a matrix such as the \textit{CatArrayBatchedCopy (Concat)}, generally exhibiting memory bound due to a large amount of data movement.

\textit{Most execution time is consumed by the kernel with reduction-tree-based computational graph.} 
This is because each resulting element of the result is computed with the reduction tree-based computational graph in all DM-Type, EW-Type, and TB-Type kernels.
Taking HAN model as an example, the \textit{sgemm} kernel occupies most execution time of \textit{Feature Projection} and \textit{Semantic Aggregation} stages. For each node, this kernel reduces the dimension of feature vector in prior stage and aggregates multiple resulting feature vectors from \textit{Neighbor Aggregation} stage into a single feature vector in later stage. 
In addition, the \textit{SpMMCsr} kernel occupies most execution time of \textit{Neighbor Aggregation} stage, which aggregates multiple feature vectors into a single vector for each node. 
Furthermore, the \textit{uEleWise} and \textit{vEleWise}, \textit{Reduce} kernels reduce multiple feature vectors or matrix into a single vector. 
The computational graphs of all these kernels are represented as a reduction tree. 
Therefore, most execution time is consumed by the kernel with reduction-tree-based computational graph in the execution of HGNNs.

\subsection{Analysis of Feature Projection Stage}\label{subsec:analysis of feature projection stage}
\begin{figure}[!t]
        \vspace{-5pt}
 	    \centering
    	\includegraphics[width = 0.8\linewidth]{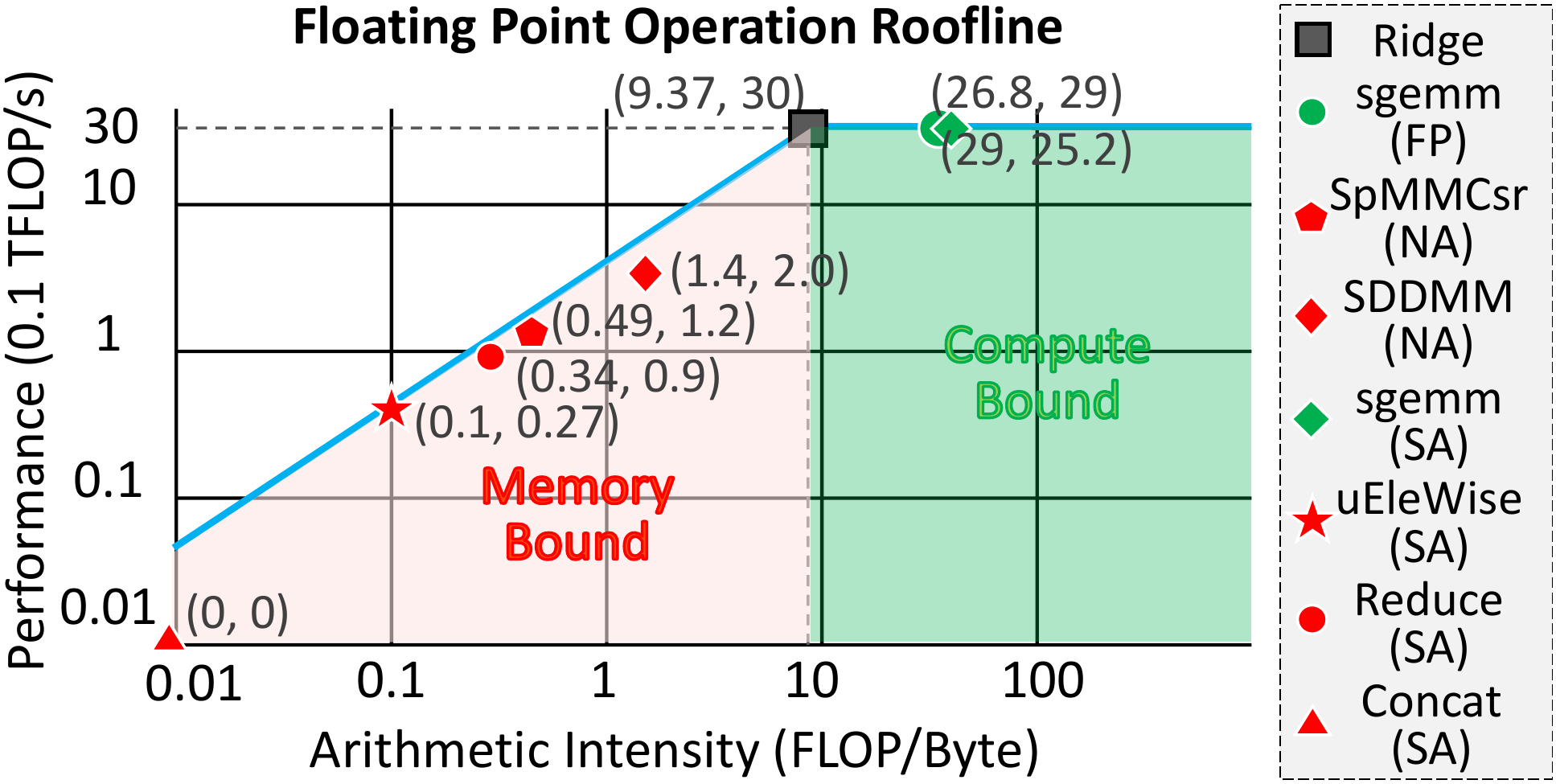}
    	\vspace{-12pt}
    	\caption{Kernels in single-precision floating point operation Roofline.}
    	\label{fig:rootline}
\end{figure}

\textit{Feature Projection stage is dominated by the execution of dense-dense matrix multiplication, primarily facing compute bound.} 
Fig. \ref{fig:stage_breakdown} shows that the DM-Type kernel (i.e., \textit{sgemm}) performing DeMM consumes most execution time of \textit{Feature Projection} stage on different HGNNs across datasets.
The \textit{sgemm} kernel exhibits high performance and high degree of data locality.
For example, the \textit{sgemm} kernel called in HAN model on DB dataset costs over 97.4\% execution time of \textit{Feature Projection} stage, as shown in Table~\ref{tab:kernel_profiling_details}. 
Due to the intensive computation of projection, this kernel achieves 95.9\% \textit{Peak Performance}. 
Due to the high reuse ratio of projection matrix, this kernel reaches 82.7\% \textit{L2 Cache Hit Rate}, 24.3\% \textit{Shared Memory Bandwidth Utilization}, and 33.6\% \textit{DRAM Bandwidth Utilization}. 
As a result, \textit{Arithmetic Intensity} of this kernel is 26.8 FLOP/Byte and larger than the one (9.37 FLOP/Byte) in the ridge of Roofline (see Fig.~\ref{fig:rootline}), which reveals that \textit{Feature Projection} stage faces compute bound.

\begin{table}[!t]
    \vspace{-10pt}
    
    \caption{Profiling results of major kernels on HAN model with DB dataset.} \label{tab:kernel_profiling_details}
    \vspace{-10pt}
    \centering
    \setlength\tabcolsep{1pt}%
	\renewcommand\arraystretch{0.8}
    \resizebox{0.4\textwidth}{!}{



\centering

\begin{tabular}{ccccccc}
\toprule
  \begin{tabular}[c]{@{}c@{}} Kernel \\ Name  \end{tabular}  & 
  \begin{tabular}[c]{@{}c@{}} Kernel \\ Type \end{tabular}   & 
  \begin{tabular}[c]{@{}c@{}} Time   \\ (\%) \end{tabular}     & 
  \begin{tabular}[c]{@{}c@{}} Peak Perf. \\ (\%) \end{tabular} & 
  \begin{tabular}[c]{@{}c@{}} DRAM BW \\ Utilization\end{tabular} & 
  \begin{tabular}[c]{@{}c@{}} Shared Memory \\ BW Utilization\end{tabular} &
  \begin{tabular}[c]{@{}c@{}} L2 Hit \\  Rate  \end{tabular} \\ \midrule \midrule
 \multicolumn{7}{c}{\blackcircledempty{2} \textit{Feature Projection}}    \\ \midrule \midrule
    sgemm   & DM  & 97.4\% & 95.9\% & 33.6\% & 24.3\% & 82.7\%   \\ \midrule \midrule
\multicolumn{7}{c}{\blackcircledempty{3} \textit{Neighbor Aggregation}}       \\ \midrule \midrule
    SpMMCsr & TB  & 85.9\% & 3.9\%  & 74.3\% & 0\%    & 31.4\% \\
    SDDMM   & TB  & 8.4\%  & 6.5\%  & 44.0\% & 0\%    & 67.6\%  \\
    \midrule \midrule
    \multicolumn{7}{c}{\blackcircledempty{4} \textit{Semantic Aggregation}}        \\ \midrule \midrule 
    sgemm   & DM  & 47.8\% & 84.2\% & 42.4\% & 21.4\% & 83.3\% \\
   uEleWise & EW  & 20\%   & 0.9\%  & 82.4\% & 0\%    & 50.0\% \\
    Reduce  & EW  & 11\%   & 3.1\%  & 88.3\% & 0\%    & 25.2\% \\  
    Concat  & DR  & 17.5\% & 0\%    & 81.6\% & 0\%    & 50.0\% \\
    \bottomrule
\end{tabular}
    }
    
    {\scriptsize `Bandwidth' and `Performance' are respectively abbreviated to BW and Perf.. The unit of arithmetic intensity is FLOP/Byte. `Peak Performance (\%)' represents the percentage of achieved performance to peak performance (FLOPS).}
  	\vspace{-10pt}
\end{table}

\subsection{Analysis of Neighbor Aggregation Stage}\label{subsec:analysis_neighbor_aggregation}
\textit{Neighbor Aggregation stage is dominated by the execution of graph-topology-based and element-wise operations, primarily facing memory bound and exhibiting irregular memory access pattern.} 
Fig. \ref{fig:stage_breakdown} shows that the TB-Type and EW-Type kernels occupy most execution time of \textit{Neighbor Aggregation} stage on different HGNNs across datasets.
Taking HAN model on DB dataset as an example, the \textit{SpMMCsr} kernel consumes over 85.9\% execution time of \textit{Neighbor Aggregation} stage, which aggregates neighboring feature vectors into a single vector for each node. 
Table~\ref{tab:kernel_profiling_details} shows that this kernel achieves high \textit{DRAM Bandwidth Utilization} (74.3\%) with low \textit{L2 Cache Hit Rate} (31.4\%). This is mainly because its memory accesses to neighboring feature vectors heavily rely on the irregular neighbor connection pattern of graph, exhibiting irregular memory access pattern. 
In addition, Fig.~\ref{fig:rootline} shows that this kernel exhibits low \textit{Arithmetic Intensity} (0.49 FLOP/Byte) and \textit{Percentage of Peak Performance} (3.9\%). 
In summary, \textit{Neighbor Aggregation} stage is mainly bounded by memory.


\subsection{Analysis of Semantic Aggregation Stage} \label{subsec:analysis_semantic_aggregation}

\textit{Semantic Aggregation stage is dominated by the execution of dense-dense matrix multiplication, element-wise operation, and data rearrangement operation, primarily facing memory bound first and then compute bound.}
In this stage, the DM-Type kernel \textit{sgemm} first calculates attention weights for each resulting feature vectors of each subgraph from \textit{Neighbor Aggregation} stage, and then the EW-type kernel \textit{uEleWise} and \textit{Reduce} aggregate these feature vectors into single one for each node with attention weights. 
The \textit{sgemm} is still compute-bound. 
On the contrary, as shown in Table~\ref{tab:kernel_profiling_details}, the \textit{uEleWise} and \textit{Reduce} kernel achieve high \textit{DRAM Bandwidth Utilization} (82.4\% and 88.3\%) with low \textit{L2 Hit Rate} (50\% and 25.2\%), on HAN model with DB dataset. 
In addition, Fig.~\ref{fig:rootline} shows that these two kernels exhibit low \textit{Arithmetic Intensity} (0.1 and 0.34 FLOP/Byte) and \textit{Percentage of Peak Performance} (0.9\% and 3.1\% ), exhibiting memory bound.
Note that RGCN, the early-stage HGNN model, directly performs \textit{Reduce} kernel, using sum operation to aggregate the resulting feature vectors without attention weights, so that its \textit{Semantic Aggregation} stage only exhibits memory bound. But attention-based aggregation is prevalent in recent, because it helps achieving better accuracy~\cite{HAN}.

\textit{The data rearrangement in \textit{Semantic Aggregation} stage is expensive.} 
A high overhead is caused by data rearrangement which aims to perform the computations of attention weights and aggregation in a manner of batch. 
For example, as shown in Table~\ref{tab:kernel_profiling_details}, the \textit{Concat} kernel, concatenating a set of vectors into a matrix, costs 17.5\% execution time of \textit{Semantic Aggregation} stage on HAN with DB dataset, exhibiting high \textit{DRAM Bandwidth Utilization} (81.6\%).


%

\subsection{Comparisons between HGNNs with GNNs}


We conduct several experiments on HAN model and a prevalent GNN (i.e., GCN~\cite{GCN}) with RD dataset to experimentally compare HGNNs with GNNs in details.

\begin{figure}
    \vspace{-5pt}
    \centering
    \includegraphics[width=0.8\linewidth]{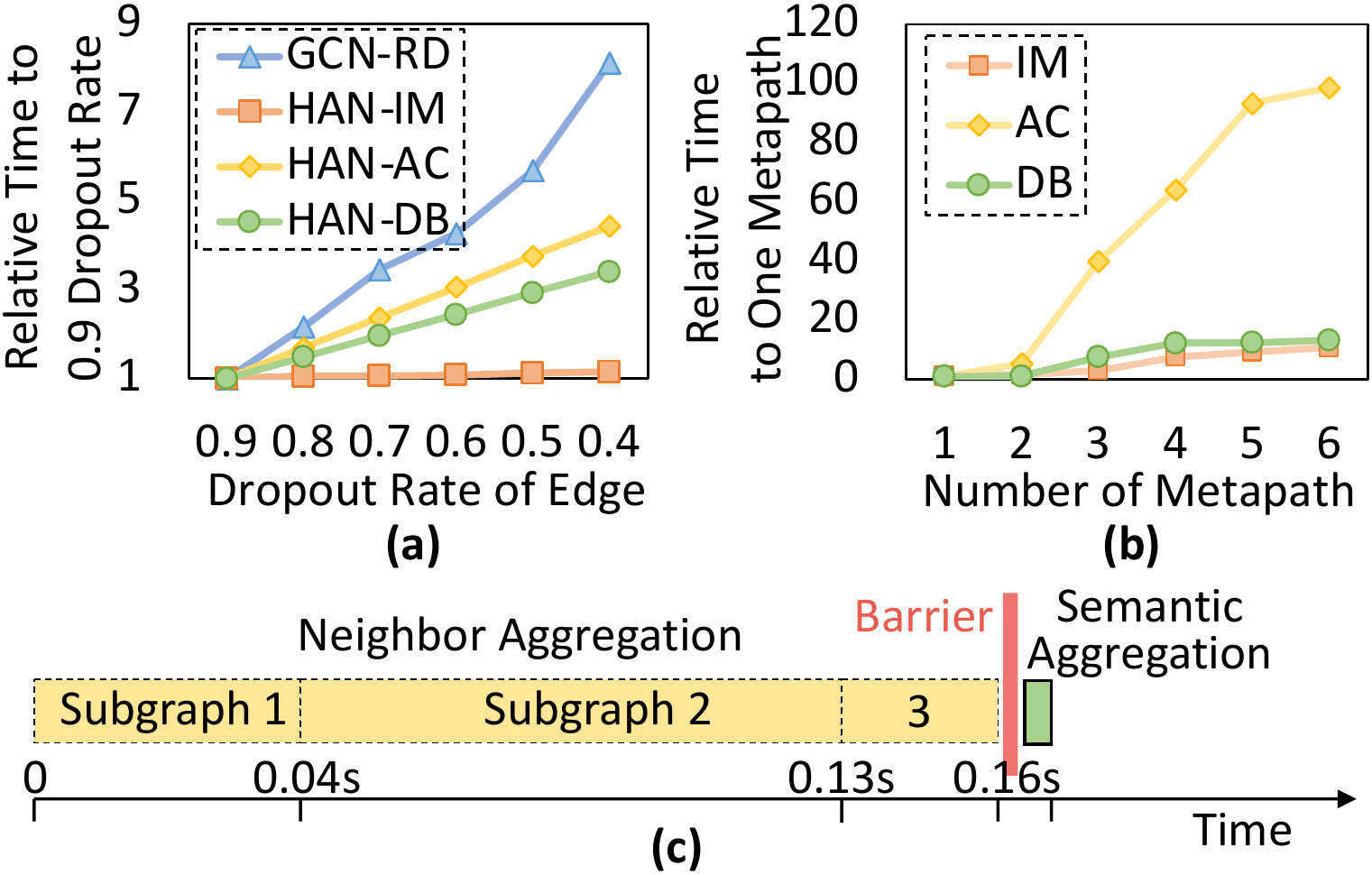}
    \vspace{-13pt}
    \caption{Comparisons: (a) Neighbor aggregation time increases as dropout rate of edge decreases (i.e., average \#neighbor increases) and (b) increases as \#metapath increases in HGNNs; (c) Timeline of \textit{Neighbor} and \textit{Semantic Aggregation} stages on HAN and DB dataset.}
    \label{fig:comparison}
    \vspace{-13pt}
\end{figure}

\textit{Except for increasing with the average number of neighbors as in \textit{Aggregation} stage in GNNs, the execution time of \textit{Neighbor Aggregation} stage in HGNNs increases further as the number of metapaths increases, as shown in Fig.~\ref{fig:comparison}(a) and (b).} 
This is because one more metapath introduces one more subgraph, which requires extra time for neighbor aggregation.


\textit{A new type of parallelism, inter-subgraph parallelism, exists in Neighbor Aggregation stage, as shown in Fig.\ref{fig:comparison}(c), except for the inter-vertex, intra-vertex, and inter-edge parallelism~\cite{gnn_characterization} as in \textit{Aggregation} stage of GNNs.}
This parallelism is derived from the independent neighbor aggregation for each subgraph. 

\textit{A barrier exists between \textit{Neighbor Aggregation} and \textit{Semantic Aggregation} stages of HGNNs as shown in Fig.\ref{fig:comparison}(c), however, one-stage aggregation of GNNs is without it}. 
This is because most HGNNs use all results of \textit{Neighbor Aggregation} stage across different subgraphs to calculate the attention weights of different semantic for the following semantic aggregation.

\subsection{Exploring The Execution of HGNN Model} \label{subsec:exploration}

\begin{figure}
    \vspace{-5pt}
    \centering
    \includegraphics[width=0.8\linewidth]{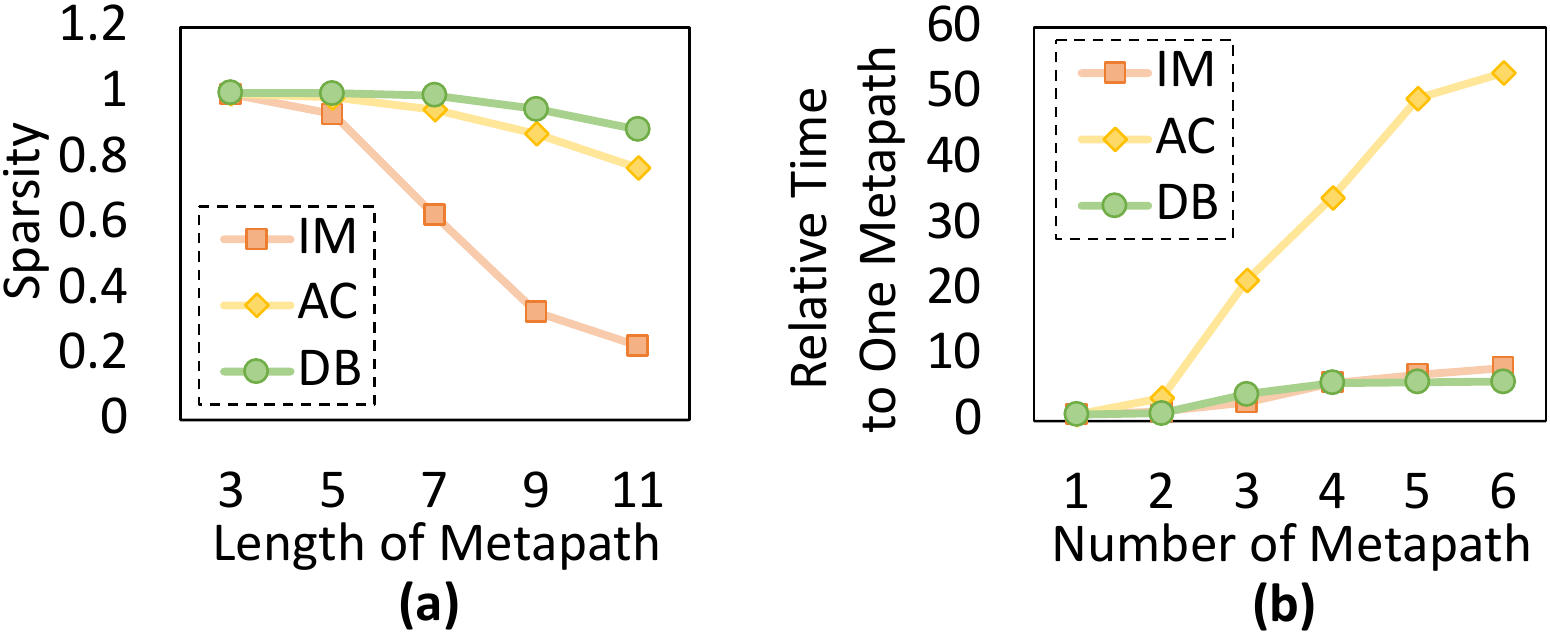}
    \vspace{-13pt}
    \caption{Exploration: (a) Sparsity of subgraph decreases as metapath length increases; (b) Execution time increases as \#metapath increases.}
    \label{fig:exploration}
    \vspace{-13pt}
\end{figure}

\textit{The sparsity of adjacency matrix for each subgraph decreases as the length of metapath increases, as shown in Fig.~\ref{fig:exploration}(a).} This is because more number of metapath-based neighbors in the same subgraph can be found for each node in \textit{Subgraph Build} stage as the length of metapath increases.

\textit{The total execution time increases as the number of metapaths increases, as shown in Fig.~\ref{fig:exploration}(b).}
This is because as the number of metapaths increases, more subgraphs are built, costing extra time to perform neighbor and semantic aggregation. 

\section{Architectural Guidelines}\label{sec:architectural guidelines}
%


From software perspective, an execution-bound-aware kernel mixing technique can be used to overlap the execution of memory-bound and compute-bound kernels to simultaneously utilize all available memory and compute resources, like Graphite~\cite{Graphite} proposed for GNN acceleration on CPU.
In addition, a subgraph-level kernel fusion technique can be used to fuse the execution of feature projection and neighbor aggregation for each subgraph to endow more opportunities for high-degree parallelism and high utilization of various resources, like fusedGCN~\cite{fuseGNN} proposed for GNN acceleration on GPU.
From hardware perspective, a correlation model can be built to quantify the relation between sparsity and the length of metapath, helping generate accurate configuration parameters for sparsity-aware optimizations and improve their effects.
In addition, a flexible dataflow scheduling equipped with a reduction-tree-based compute unit can be used to exploit the parallelism and reduce data movements among the reduction-tree-based computational graph for the efficient inference. 

\section{Conclusions}\label{sec:conclusions}


In this work, we characterize and explore an emerging application HGNNs on GPU.
We believe our observations and conclusions will help related researchers understand the execution pattern and execution semantic of HGNNs.

\ifCLASSOPTIONcaptionsoff
  \newpage
\fi

\bibliographystyle{plain}
\bibliography{ref.bib}


\end{document}